\documentclass[a4paper,12pt]{article}

\usepackage{graphicx}
\usepackage{epsfig}
\usepackage{amsfonts,amsmath}
\usepackage{amssymb}
\usepackage{epsf}
\usepackage{hyperref}
\usepackage{authblk}
\usepackage{cite}
\usepackage{placeins}
\usepackage[dvipsnames]{xcolor}

\usepackage{ulem}
\newcommand{\jk}[1]{{\textcolor{black}{{#1}}}}
\newcommand{\bk}[1]{{\textcolor{black}{{#1}}}}

\title{Static multipolar Einstein-vector-Gauss-Bonnet black holes}

\author[1]{Burkhard Kleihaus
\thanks{\href{mailto:b.kleihaus@uni-oldenburg.de}{b.kleihaus@uni-oldenburg.de}}}
\author[1]{Jutta Kunz
\thanks{\href{mailto:jutta.kunz@uni-oldenburg.de}{jutta.kunz@uni-oldenburg.de}}}

\affil[1]{Institut f\"ur Physik, Universit\"at Oldenburg, Postfach 2503,
D-26111 Oldenburg, Germany}

\date{\today}

\begin{document}

\maketitle

\begin{abstract}
We construct and analyze static multipolar black holes in Einstein-vector-Gauss-Bonnet theory with a quadratic coupling function.  
The vectorized solutions bifurcate from the Schwarzschild black hole at discrete values of the Gauss-Bonnet coupling, obtained from a perturbative eigenvalue problem on the Schwarzschild background and labelled by the angular multipole number $\ell$.
We restrict to the fundamental radial branches.  
The electric sector contains the known spherically symmetric $\ell=0$ branch as well as axisymmetric branches with $\ell>0$.  
These electric branches extend to larger values of the coupling.
The magnetic sector features a sequence of axisymmetric branches, including the previously found magnetic dipole branch at $\ell=0$.  
The magnetic branches instead exist only on finite intervals of the coupling, ending at critical solutions. 
Away from the bifurcation point, nonlinearities generate additional multipoles, but even-$\ell$ and odd-$\ell$ moments remain separated.
%reflection symmetry.
\end{abstract}

\section{Introduction}

General Relativity (GR) may be regarded as the leading term in a low-energy effective field theory expansion of gravitational dynamics, organized in powers of derivatives and curvature \cite{Donoghue:1995cz,Burgess:2003jk}.
This point of view allows possible deviations from GR to be parametrized in different regimes, from cosmological evolution to strong-field compact objects \cite{Berti:2015itd,CANTATA:2021ktz}.
Extensions of GR with additional scalar or vector degrees of freedom arise in dark-energy and dark-matter models, low-energy limits of more fundamental theories, and beyond-standard-model physics \cite{Faraoni:2010pgm,CANTATA:2021ktz}.
Such theories can modify cosmological dynamics and strong-field gravity while recovering GR in regimes where it is well tested.

One class of alternative gravity theories couples additional fields to higher-curvature invariants. 
In the scalar sector, Horndeski theory provides the most general scalar-tensor framework with second-order field equations~\cite{Horndeski:1974wa,Charmousis:2011bf,Kobayashi:2011nu}.
Much work has focussed on Einstein-scalar-Gauss-Bonnet (EsGB) theories, which also arise in low-energy effective string theories, see, e.g., \cite{Moura:2006pz}.
EsGB theories admit scalarized compact objects such as black holes and neutron stars, whose properties depend strongly on the coupling function of the scalar field to the Gauss-Bonnet (GB) invariant.

In dilatonic and shift-symmetric EsGB theories with exponential and linear coupling functions, respectively, stationary black holes are always scalarized, and GR black holes are no longer solutions of the field equations \cite{Kanti:1995vq,Torii:1996yi,Pani:2009wy,Kleihaus:2011tg,Sotiriou:2014pfa}.
Employing $Z_2$ symmetric coupling functions, on the other hand, allows for the GR black holes and, in addition, scalarized black holes \cite{Doneva:2017bvd,Silva:2017uqg,Antoniou:2017acq}.
Spontaneous scalarization was first identified for compact stars in scalar-tensor theory, where sufficiently compact neutron stars can develop a nontrivial scalar field even when the weak-field regime remains close to GR~\cite{Damour:1993hw}.

For black holes, an analogous mechanism occurs in Einstein-scalar-Gauss-Bonnet (EsGB) theory where the scalarization is induced by curvature.
When the scalar field couples appropriately to the Gauss-Bonnet invariant, GR black holes may become tachyonically unstable and bifurcate to develop scalar hair \cite{Doneva:2017bvd,Silva:2017uqg,Antoniou:2017acq,Cunha:2019dwb,Collodel:2019kkx}.
Scalarization can also be induced by sufficiently strong rotation \cite{Dima:2020yac,Herdeiro:2020wei,Berti:2020kgk,Hod:2020jjy,Doneva:2020nbb}.
In addition to the usual tachyonic mechanism, nonlinear effects can produce further families of scalarized black holes \cite{Doneva:2021tvn,Blazquez-Salcedo:2022omw,Doneva:2022yqu,Lai:2023gwe}.
A recent overview of scalarized black holes is given in Ref.~\cite{Doneva:2022ewd}.

Generalized Proca theories extend these ideas to vector degrees of freedom while preserving second-order field equations \cite{Horndeski:1976gi,Tasinato:2014eka,Heisenberg:2014rta,Tasinato:2014mia}.
Black-hole solutions in vector-tensor theories have been investigated in generalized Proca theories and related models \cite{Chagoya:2016aar,Fan:2016jnz,Babichev:2017rti,Chagoya:2017fyl,Heisenberg:2017xda,Heisenberg:2017hwb,Verbin:2020fzk,Oliveira:2020dru,Barton:2021wfj,Minamitsuji:2024ygi,Charmousis:2025jpx,Eichhorn:2025pgy,Konoplya:2025uiq,Lutfuoglu:2025qkt,Konoplya:2025bte,Fernandes:2026rjs,Kleihaus:2026rev}.
Analogously to spontaneous scalarization, vector-tensor theories may also exhibit spontaneous vectorization, where a tachyonic instability gives rise to a new branch of compact-object or black-hole solutions with nontrivial vector hair \cite{Ramazanoglu:2017xbl,Ramazanoglu:2018tig,Ramazanoglu:2019gbz,Ramazanoglu:2019jrr}.
In Einstein-vector-Gauss-Bonnet (EvGB) theory this mechanism arises from the coupling of the vector field to the Gauss-Bonnet term through a suitable function of $A_\mu A^\mu$.

For the quadratic EvGB model considered here, static spherically symmetric vectorized black holes were first constructed in Ref.~\cite{Barton:2021wfj}.
They carry an electric vector charge and bifurcate from the Schwarzschild solution at the onset of a tachyonic instability.  
Ref.~\cite{Kleihaus:2026rev} extended this solution space to include a static axially symmetric magnetic branch with a magnetic dipole moment, as well as radial excitations and rotating generalizations. 
In particular, it was found that rotation connects the electric and magnetic domains of solutions.
Here we keep the spacetime strictly static and classify the vectorized EvGB black holes according to their angular multipolar structure.
We denote the angular multipole number by $\ell$.
Following Ref.~\cite{Kleihaus:2026rev}, we denote the radial excitation number by $n$.
In the numerical study we then restrict to the fundamental radial branches, i.e., $n=0$, and therefore suppress $n$ in most expressions.

We find a sequence of static multipolar vectorized black holes in each sector.
For the fundamental radial sector, we construct the branches with $\ell=0,1,2,3$. 
Although the electric and magnetic sectors have analogous multipolar bifurcations, their nonlinear continuations differ.

Static nonspherical vacuum black holes are excluded in GR by Israel's theorem \cite{Israel:1967wq}.  
The solutions constructed here evade this conclusion because the vector field is nontrivially coupled to the Gauss-Bonnet invariant.
They therefore constitute examples of static, asymptotically flat, axisymmetric black holes without  rotation.
Static nonspherical black holes can also arise in GR in the presence of suitable matter content, such as non-Abelian fields \cite{Kleihaus:1997ic}.

The paper is organized as follows.  
Section~\ref{sec:theory} presents the action, field equations, ansatz, boundary conditions, and global quantities.  
In Section~\ref{sec:results} we first derive the perturbative bifurcation problem and the discrete multipolar eigenvalues.  
We then present the nonlinear electric and magnetic branches and compare their domains of existence and multipole structure.  
Section~\ref{sec:conclusions} contains the conclusions and outlook.

\section{Theoretical setting}\label{sec:theory}

\subsection{Action and equations of motion}

We start with the effective EvGB action
\begin{eqnarray}
S=\frac{1}{16 \pi}\int \left[R - F_{\mu\nu}F^{\mu\nu}
 + \lambda A_\mu A^\mu R^2_{\rm GB}
 \right] \sqrt{-g} d^4x  \ ,
\label{act}
\end{eqnarray}
where $R$ denotes the curvature scalar, $F_{\mu\nu}=\partial_\mu A_\nu-
\partial_\nu A_\mu$ is the field strength tensor of the vector field, and
\begin{eqnarray}
R^2_{\rm GB} = R_{\mu\nu\rho\sigma} R^{\mu\nu\rho\sigma}
- 4 R_{\mu\nu} R^{\mu\nu} + R^2
\end{eqnarray}
is the Gauss-Bonnet invariant.  
The vector field couples to the Gauss-Bonnet term via the quadratic coupling function $\lambda A_\mu A^\mu$, with coupling parameter $\lambda$.  
Although $R^2_{\rm GB}$ is topological in four dimensions, it contributes nontrivially to the field equations when coupled to the vector field.

Variation of the action with respect to the vector field and the metric yields
\begin{equation}
\nabla_\mu F^{\mu\nu} = -\frac{\lambda}{2} A^\nu R^2_{\rm GB} \ ,
\label{scleq}
\end{equation}
and
\begin{equation}
G_{\mu\nu}  = \frac{1}{2}T^{({\rm eff})}_{\mu\nu} \ .
\label{Einsteq}
\end{equation}
The effective stress-energy tensor is
\begin{equation}
T^{({\rm eff})}_{\mu\nu} = T^{(A)}_{\mu\nu} -2 T^{(GB)}_{\mu\nu} \ ,
\label{teff}
\end{equation}
with
\begin{equation}
T^{(A)}_{\mu\nu} = 4 F_\mu^{\phantom{\mu}\kappa} F_{\nu\kappa}
            -g_{\mu\nu} \left(F_{\rho\kappa}F^{\rho\kappa}\right)\ ,
\label{tphi}
\end{equation}
and
\begin{equation}
T^{(GB)}_{\mu\nu} =
\frac{\lambda}{2}\left(g_{\rho\mu} g_{\beta\nu}+g_{\beta\mu} g_{\rho\nu}\right)
\eta^{\kappa\beta\alpha\delta}\tilde{R}^{\rho\gamma}_{\phantom{\rho\gamma}\alpha\delta}
\nabla_\gamma \nabla_\kappa \left(A_\sigma A^\sigma\right) +
\lambda A_\mu A_\nu\, R^2_{\rm GB} \ ,
\label{teffi}
\end{equation}
where $\tilde{R}^{\rho\gamma}_{\phantom{\rho\gamma}\alpha\delta}=
\eta^{\rho\gamma\xi\tau}R_{\xi\tau\alpha\delta}$ and
$\eta^{\rho\gamma\xi\tau}=\epsilon^{\rho\gamma\xi\tau}/\sqrt{-g}$.

To obtain static axially symmetric black holes, we employ isotropic coordinates for the line element
\begin{equation}
ds^2 = -f_0 dt^2 + f_1\left[ f_2 \left(dr^2+r^2 d\theta^2\right)
                            +r^2\sin^2\theta d\varphi^2 \right]\ ,
\label{met}
\end{equation}
and take the vector field to be either purely electric,
\begin{equation}
A_\mu dx^\mu = A_t dt \ ,
\label{vec_el}
\end{equation}
or purely magnetic,
\begin{equation}
A_\mu dx^\mu = A_\varphi d\varphi \ .
\label{vec_mg}
\end{equation}
All functions depend only on $r$ and $\theta$.  
The terms ``electric'' and ``magnetic'' are used only as convenient labels.  
The vector field is not an electromagnetic gauge potential, since the coupling function $\lambda A_\mu A^\mu$ breaks the gauge invariance.

Within a strictly static ansatz, solutions with both $A_t\neq0$ and
$A_\varphi\neq0$ are not consistent.  
In that case the stress-energy component $T_{t\varphi}$ is proportional to $A_t A_\varphi$ and represents a source for angular momentum.
Static black holes are therefore separated into distinct electric and magnetic sectors.

The EvGB equations reduce to one PDE for the nonzero vector-field function and
five PDEs for the metric functions.  
We introduce the compactified coordinate $x=1-r_{\rm H}/r$, which maps the horizon to $x=0$ and spatial infinity to $x=1$.  
We extract the double zero of $f_0$ at the horizon by redefining $f_0\to x^2 f_0$ and write $A_t=x^2 V$.  
For the magnetic sector we use $A_\varphi=H\sin^2\theta$.  
The coupling function then takes the form
\begin{equation}
\lambda A_\mu A^\mu
      = \lambda \left(\frac{H^2 \sin^2\theta}{r^2 f_1} - \frac{x^2 V^2}{f_0} \right) \ .
\label{coupA}
\end{equation}

\subsection{Boundary conditions}

Regularity at the horizon requires
\begin{equation}
  f_0-\partial_x f_0 = 0 \ , \quad
2 f_1+\partial_x f_1 = 0 \ , \quad
\partial_x f_2 = 0 \ , \quad
 V-\partial_x V = 0 \ , \quad
\partial_x H = 0 \ .
\label{BCrh}
\end{equation}
\jk{The horizon boundary conditions follow from regular near-horizon expansions in the compactified coordinate $x$. With the double zero of the lapse extracted, a regular nonextremal horizon is described by
\begin{equation}
f_0=f_{00}+f_{00}x+O(x^2),\quad
f_1=f_{10}-2f_{10}x+O(x^2),\quad
f_2=f_{20}+O(x^2).
\end{equation}
For the electric sector we write $A_t=x^2V$, so that the vector potential vanishes at the horizon and the invariants $A_\mu A^\mu$ and $F_{\mu\nu}F^{\mu\nu}$ remain finite. 
The vector equation then gives
\begin{equation}
V=V_0+V_0x+O(x^2),
\end{equation}
which yields $V-\partial_xV=0$ at the horizon. In the magnetic sector regularity gives an even expansion,
\begin{equation}
H=H_0+O(x^2),
\end{equation}
and hence $\partial_xH=0$.}

At spatial infinity we impose asymptotic flatness and a vanishing vector field,
\begin{equation}
f_i = 1 \ , \quad i=0,1,2 \ , \qquad
V = 0 \ , \qquad H = 0 \ .
\label{BCinf}
\end{equation}
On the symmetry axis, elementary flatness and regularity give
\begin{equation}
\partial_\theta f_0 = 0 \ , \quad
\partial_\theta f_1 = 0 \ , \quad
f_2 = 1 \ , \quad
\partial_\theta V = 0 \ , \quad
\partial_\theta H = 0 \ .
\label{BCz}
\end{equation}
On the equatorial plane we demand reflection symmetry of the metric,
\begin{equation}
\partial_\theta f_i = 0 \ , \qquad i=0,1,2 \ ,
\label{BCeqmet}
\end{equation}
%
%The vector-field functions may be either symmetric or antisymmetric,
whereas the vector-field functions may be either symmetric or antisymmetric,
\begin{eqnarray}
\partial_\theta V = 0 \ , \quad \partial_\theta H = 0
      && ({\rm symmetric})\ ,
\label{BCeqvec1}\\
V = 0 \ , \quad H = 0
      && ({\rm antisymmetric})\ .
\label{BCeqvec2}
\end{eqnarray}

\subsection{Physical properties}

The ADM mass $M$ and the electric and magnetic multipole moments are obtained from the asymptotic behavior
\begin{equation}
x^2 f_0 \to 1-\frac{2 M}{r} \ , \qquad
V(r,\theta)  \to \sum_{\ell=0}^{\infty} q_\ell
       \frac{P_\ell(\cos\theta)}{r^{\ell+1}} \ , \qquad
H(r,\theta)  \to \sum_{\ell=0}^{\infty} \mu_\ell
       \frac{\tilde{P}_\ell(\cos\theta)}{r^{\ell+1}} \ .
\label{MJinf}
\end{equation}
The polynomials $P_\ell$ and $\tilde{P}_\ell$ are defined in Section~\ref{PerturbativeSolutions}.  
The monopole $q_0$ is the electric vector charge, while $\mu_0$ gives the magnetic dipole moment with the present normalization of $A_\varphi$.

The metric on a spatial cross-section of the horizon $\Sigma_{\rm H}$ is
\begin{eqnarray}
\label{horizon-metric}
d\Sigma^2_{\rm H}=h_{ij} dx^i dx^j=
r_{\rm H}^2 f_1(r_{\rm H},\theta )
\left(  f_2(r_{\rm H},\theta ) d\theta^2 + \sin^2\theta d\varphi^2\right) \ .
\end{eqnarray}
The horizon area is
\begin{eqnarray}
\label{AH}
A_{\rm H}=4\pi r_{\rm H}^2 \int_0^{\pi/2} f_1\sqrt{f_2}\sin\theta d\theta \ .
\end{eqnarray}
\bk{The polar and equatorial horizon radii are
\begin{eqnarray}
\label{RpRe}
R_p=  r_{\rm H} \frac{2}{\pi} \int_0^{\pi/2}\sqrt{f_1 f_2} d\theta \ , \qquad
R_e =  r_{\rm H} \left. \sqrt{f_1} \right|_{\theta=\pi/2} \ .
\end{eqnarray}
}

The entropy is obtained from the Wald-Iyer Noether-charge expression
\cite{Wald:1984rg,Lee:1990nz,Wald:1993nt,Iyer:1994ys} (see also \cite{Hajian:2015xlp,Ghodrati:2016vvf,Hajian:2020dcq})
\begin{eqnarray}
\label{S-Noether}
S_{\rm H}=
\frac{1}{4}\int_{\Sigma_{\rm H}}
\sqrt{h}\left[1+ 2 \lambda A_\mu A^\mu \tilde R \right] d^{2}x
= \pi r_{\rm H}^2\int_0^{\pi/2}
\left[1+ 2 \lambda \frac{\sin^2\theta H^2}{r_{\rm H}^2 f_1} \tilde R \right]
f_1\sqrt{f_2}\sin\theta d\theta \ .
\end{eqnarray}
Here $h$ is the determinant of the horizon metric and $\tilde R$ is the scalar
curvature of the horizon.  
For electric EvGB black holes, the Gauss-Bonnet contribution to the entropy vanishes at the horizon because $A_t=x^2V$.

The Hawking temperature is
\jk{ 
\begin{equation}
 T_H =
\frac{1}{2\pi r_H}
\left.
\sqrt{\frac{f_0}{f_1 f_2}}
\right|_{r=r_H} .   
\end{equation}
For regular solutions this expression is independent of $\theta$ on the horizon, as required by the zeroth law.
}
%\begin{eqnarray}
%\label{TH}
%T_{\rm H}=\frac{1}{2 \pi r_{\rm H}} \sqrt{\frac{f_0}{f_1 f_2}} \ .
%\end{eqnarray}
%
%and the free energy is
%
%\begin{eqnarray}
%\label{F}
%F= M - T_{\rm H} S_{\rm H} \ .
%\end{eqnarray}
%

\jk{For variations at fixed coupling $\lambda$ and with the asymptotic vector field set to zero, the vector multipole moments are not independent conserved charges. 
Thus no separate vector work term is expected, and the corresponding first-law variation reduces to
\begin{equation}
\delta M = T_H \delta S_H .
\end{equation}
The numerical branches displayed below, however, are generated by varying the dimensionless coupling $\lambda/M^2$ at fixed horizon coordinate $r_H$.}
\jk{We therefore use $S_H$ and $T_H$ mainly to characterize the solutions along the branches and do not perform a full thermodynamic analysis.
Nevertheless, as a numerical consistency check, we have verified the first-law relation in scaled form. 
With
%\begin{equation}
%\alpha=\frac{\lambda}{M^2}, \qquad
%\bar S=\frac{S_H}{4\pi M^2}, \qquad
%\bar T=8\pi T_H M ,
%\end{equation}
\bk{
\begin{equation}
\bar \lambda =\frac{\lambda}{M^2}, \qquad
\bar S=\frac{S_H}{4\pi M^2}, \qquad
\bar T=8\pi T_H M ,
\end{equation}
}
the first law at fixed $\lambda$ becomes
%\begin{equation}
%\bar T\left(\frac{d\bar S}{d\alpha}-\frac{\bar S}{\alpha}\right)
%= -\frac{1}{\alpha}.
%\end{equation}
\bk{
\begin{equation}
\bar T\left(\frac{d\bar S}{d\bar\lambda}-\frac{\bar S}{\bar\lambda}\right)
= -\frac{1}{\bar\lambda}.
\end{equation}
}
The residual of this relation remains small along the numerical branches, within the accuracy expected from the finite-difference evaluation of % $d\bar S/d\alpha$.}
\bk{$d\bar S/d\bar \lambda$}
}
%The polar and equatorial horizon radii are
%
%\begin{eqnarray}
%\label{RpRe}
%R_p=  r_{\rm H} \frac{2}{\pi} \int_0^{\pi/2}\sqrt{f_1 f_2} d\theta \ , \qquad
%R_e =  r_{\rm H} \left. \sqrt{f_1} \right|_{\theta=\pi/2} \ .
%\end{eqnarray}

\section{Results}\label{sec:results}

\subsection{Numerical construction}

The nonlinear solutions are obtained by solving the elliptic boundary-value
problem for the metric and vector-field functions on the compactified domain
$0\leq x\leq1$, $0\leq\theta\leq\pi/2$.  
The numerical construction follows the same procedure as the one used for the stationary solutions of Ref.~\cite{Kleihaus:2026rev}.  
The horizon coordinate $r_{\rm H}$ is kept fixed, while the coupling parameter
$\lambda$ is varied along a branch.  
The numerical branches are characterized by dimensionless quantities that are scaled by the ADM mass $M$.  
The critical endpoints are identified by continuing a branch up to the point where the nonlinear solver ceases to converge to regular black hole solutions.

\subsection{Perturbative multipolar bifurcations}\label{PerturbativeSolutions}

The couplings at which vectorized black holes bifurcate from the Schwarzschild solution are determined at lowest order by the vector equation on the Schwarzschild background.  
In the perturbative calculations the angular and radial dependences separate. 
The angular multipole number is denoted by $\ell$. 
The radial excitation number $n$ counts the nodes of the function $\gamma_{\ell n}(r)$. 
In the following we restrict to the fundamental radial branch, $n=0$, and write
simply $\gamma_\ell(r)$ and $\lambda_\ell$.

For the electric perturbations we write
\begin{equation}
V=\gamma_\ell(r) P_\ell(\cos\theta) \ ,
\end{equation}
where $P_\ell$ are Legendre polynomials.  
The radial function obeys
\begin{equation}
\gamma_\ell^{''}
      +\frac{2}{r}\frac{2+ (r/r_{\rm H})^2}{(r/r_{\rm H})^2-1}\, \gamma_\ell'
      -\frac{1}{r^2}\left[\frac{6}{(r/r_{\rm H})^2-1}+\ell(\ell+1)\right] \gamma_\ell
      +\frac{96 \lambda_\ell}{r_{\rm H}^4}\frac{(r/r_{\rm H})^2}{(1+r/r_{\rm H})^8} \gamma_\ell
 = 0 \ ,
\label{pert_el_ode}
\end{equation}
with boundary conditions $r_{\rm H}\gamma_\ell'(r_{\rm H})-\gamma_\ell(r_{\rm H})=0$
and $\gamma_\ell(\infty)=0$.  
In the asymptotic region, $\gamma_\ell(r)\sim r^{-(\ell+1)}$.

\begin{figure}[t!]
\begin{center}
%\mbox{
%(a)\hspace*{-0.5cm}\includegraphics[height=.22\textheight, angle =0]{gamma_el_vs_x.eps}
%(b)\hspace*{-0.5cm}\includegraphics[height=.22\textheight, angle =0]{P_el_vs_th.eps}
%}
\mbox{
(a)\hspace*{-0.5cm}\includegraphics[height=.22\textheight, angle =0]{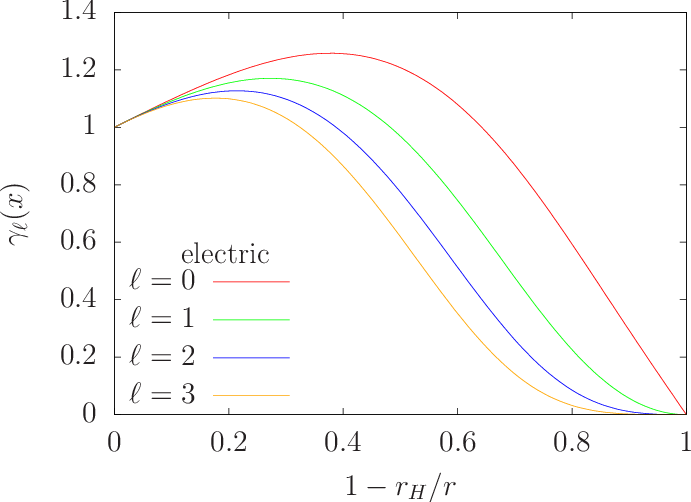}
(b)\hspace*{-0.5cm}\includegraphics[height=.22\textheight, angle =0]{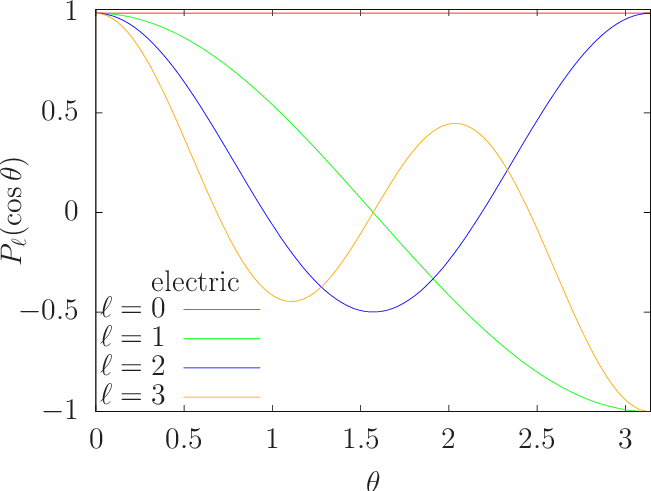}
}
\end{center}
\caption{
\jk{Electric vector perturbations on the Schwarzschild background at
the bifurcation points: (a) radial eigenfunctions $\gamma_\ell$ and
(b) Legendre polynomials $P_\ell(\cos\theta)$ for $\ell=0,\ldots,3$.}}
%Perturbative electric EvGB black holes: (a) radial functions
%$\gamma_\ell$ and (b) Legendre polynomials $P_\ell(\cos\theta)$ for
%$\ell=0,\ldots,3$.}
\label{fig:fig1}
\end{figure}

For magnetic perturbations we write
\begin{equation}
H=\gamma_\ell(r) \tilde{P}_\ell(\cos\theta) \ .
\end{equation}
The polynomials $\tilde{P}_\ell(y)$ are related to Legendre polynomials by
\begin{equation}
\tilde{P}_\ell(y) = \frac{2}{(\ell+1)(\ell+2)}\frac{d}{dy}P_{\ell+1}(y) \ ,
\label{tildP}
\end{equation}
with normalization $\tilde{P}_\ell(1)=1$.  
The radial function satisfies
\begin{equation}
\gamma_\ell^{''}
      -\frac{2}{r}\frac{1-2 r/r_{\rm H}}{(r/r_{\rm H})^2-1}\, \gamma_\ell'
      -\frac{2+\ell(\ell+3)}{r^2}\, \gamma_\ell
      +\frac{96 \lambda_\ell}{r_{\rm H}^4}\frac{(r/r_{\rm H})^2}{(1+r/r_{\rm H})^8} \gamma_\ell
 = 0 \ ,
\label{pert_mg_ode}
\end{equation}
with boundary conditions $\gamma'_\ell(r_{\rm H})=0$ and $\gamma_\ell(\infty)=0$.
Again, $\gamma_\ell(r)\sim r^{-(\ell+1)}$ at infinity.

\begin{figure}[t!]
\begin{center}
%\mbox{
%(a)\hspace*{-0.5cm}\includegraphics[height=.22\textheight, angle =0]{gamma_mg_vs_x.eps}
%(b)\hspace*{-0.5cm}\includegraphics[height=.22\textheight, %angle =0]{P_mg_vs_th.eps}
%}
\mbox{
(a)\hspace*{-0.5cm}\includegraphics[height=.25\textheight, angle =0]{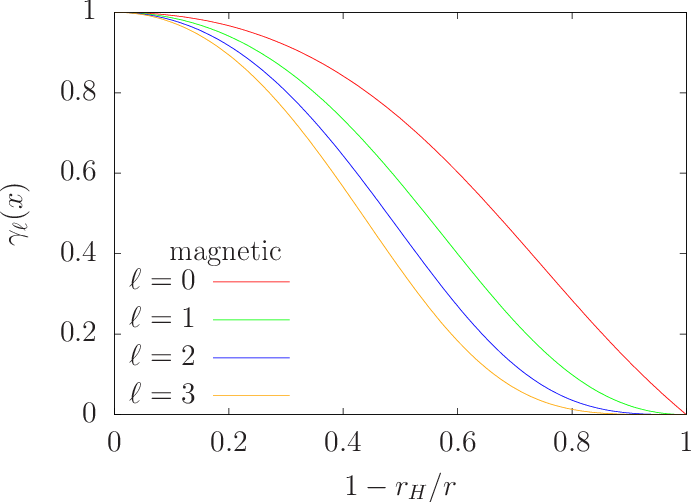}
(b)\hspace*{-0.5cm}\includegraphics[height=.25\textheight, angle =0]{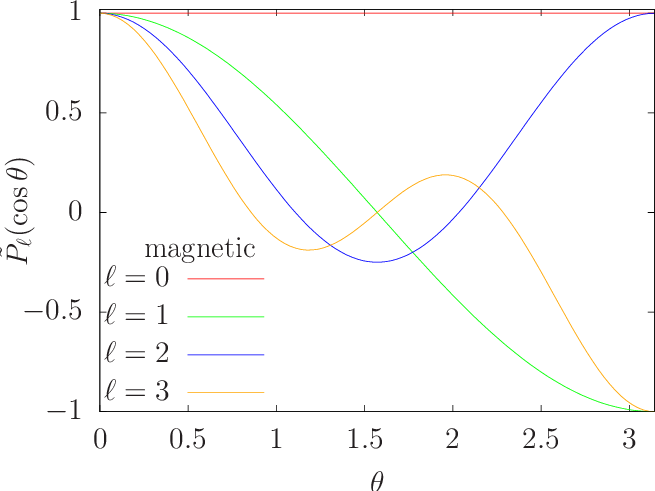}
}
\end{center}
\caption{
\jk{Magnetic vector perturbations on the Schwarzschild background at the bifurcation points: (a) radial eigenfunctions $\gamma_\ell$ and (b) polynomials $\tilde P_\ell(\cos\theta)$ for $\ell=0,\ldots,3$.}}
%Perturbative magnetic EvGB black holes: (a) radial functions
%$\gamma_\ell$ and (b) polynomials $\tilde{P}_\ell(\cos\theta)$ for
%$\ell=0,\ldots,3$.}
\label{fig:fig2}
\end{figure}

For a fixed angular index $\ell$, Eqs.~(\ref{pert_el_ode}) and (\ref{pert_mg_ode}) possess nontrivial solutions only for discrete values of the coupling.  
The first values of $\lambda_\ell/M^2$ for the fundamental radial branch are given in Table~\ref{tab:lambda}.  
The corresponding radial functions and angular polynomials are displayed in Figs.~\ref{fig:fig1} and \ref{fig:fig2}.

\begin{table}[h!]
\begin{center}
\begin{tabular}{|c|c|c|}
\hline
$\ell$ & electric & magnetic  \\
\hline
0 &  4.681747 &  3.175894 \\
1 &  8.612875 &  7.297495 \\
2 & 13.915255 & 12.685844 \\
3 & 20.564275 & 19.383946 \\
\hline
\end{tabular}
\caption{Bifurcation values $\lambda_\ell/M^2$ for $\ell=0,\ldots,3$ on the
fundamental radial branch.}
\label{tab:lambda}
\end{center}
\end{table}

\FloatBarrier

\subsection{Nonlinear electric branches}\label{sec:electric}

Electric black holes are characterized by $A_t\neq0$ and $A_\varphi=0$.  
They bifurcate from the Schwarzschild black holes at the electric values $\lambda_\ell$ in Table~\ref{tab:lambda}.  
The $\ell=0$ branch is the known spherically symmetric branch of Ref.~\cite{Barton:2021wfj}.  
The branches with $\ell>0$ are genuinely axisymmetric.  
In all cases considered here the electric branches continue to larger values of $\lambda/M^2$; no upper endpoint was found.

\begin{figure}[p!]
\begin{center}
\mbox{
(a)\hspace*{-0.5cm}\includegraphics[height=.22\textheight, angle =0]{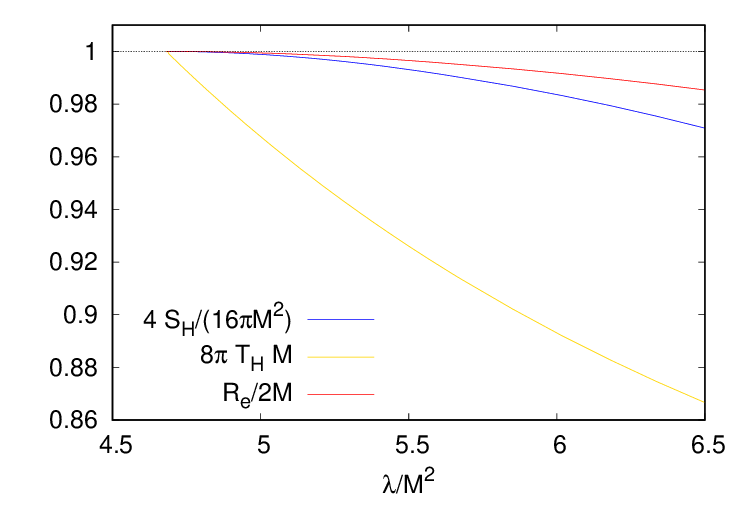}
(b)\hspace*{-0.5cm}\includegraphics[height=.22\textheight, angle =0]{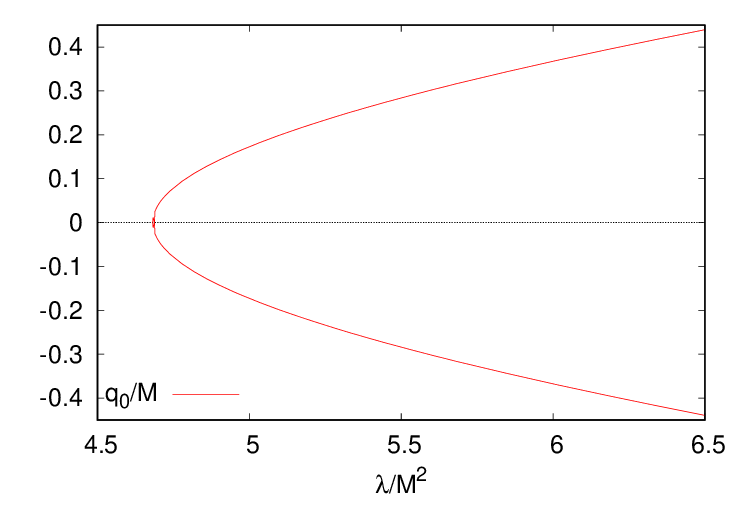}
}
\mbox{
(c)\hspace*{-0.5cm}\includegraphics[height=.22\textheight, angle =0]{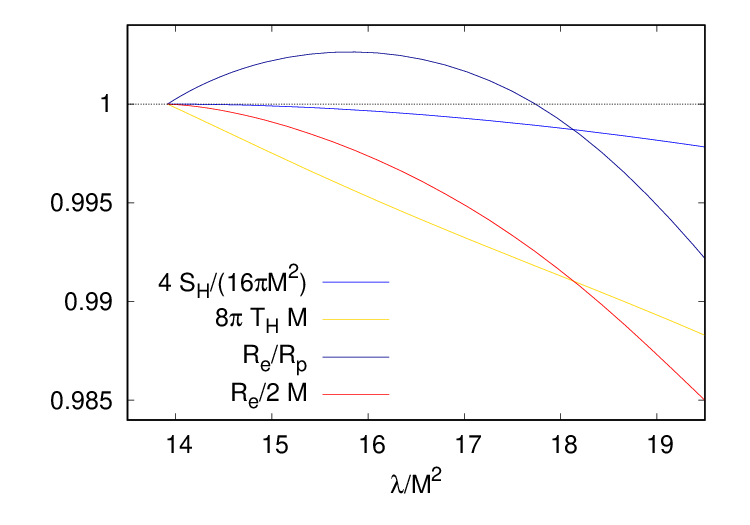}
(d)\hspace*{-0.5cm}\includegraphics[height=.22\textheight, angle =0]{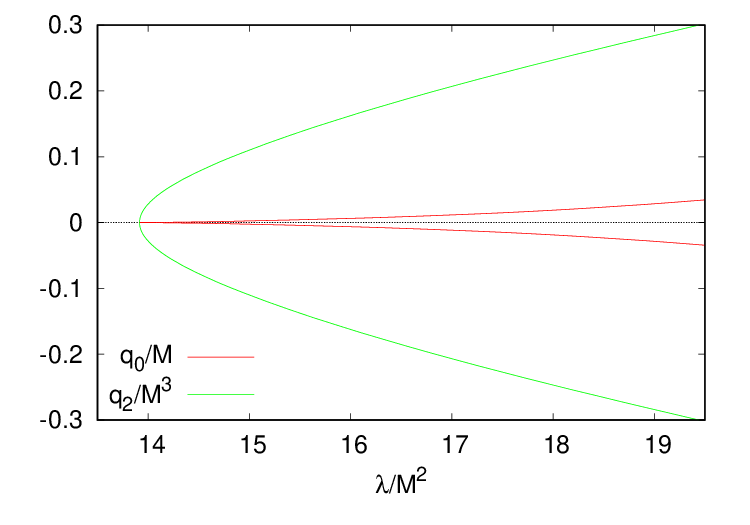}
}
\mbox{
(e)\hspace*{-0.5cm}\includegraphics[height=.22\textheight, angle =0]{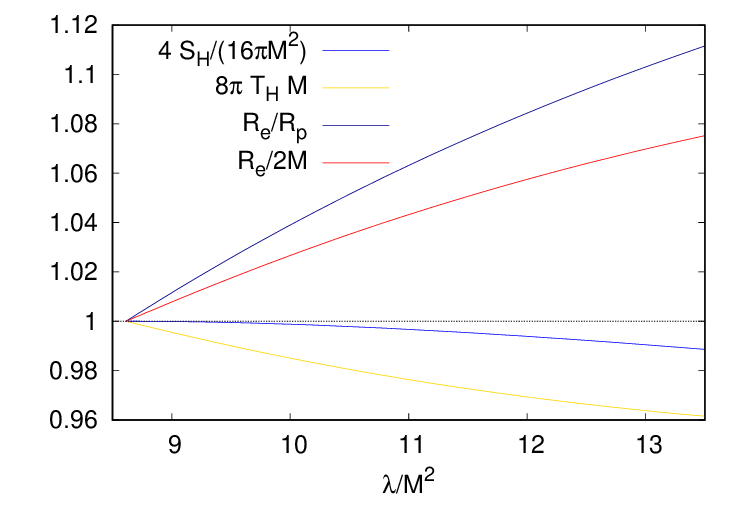}
(f)\hspace*{-0.5cm}\includegraphics[height=.22\textheight, angle =0]{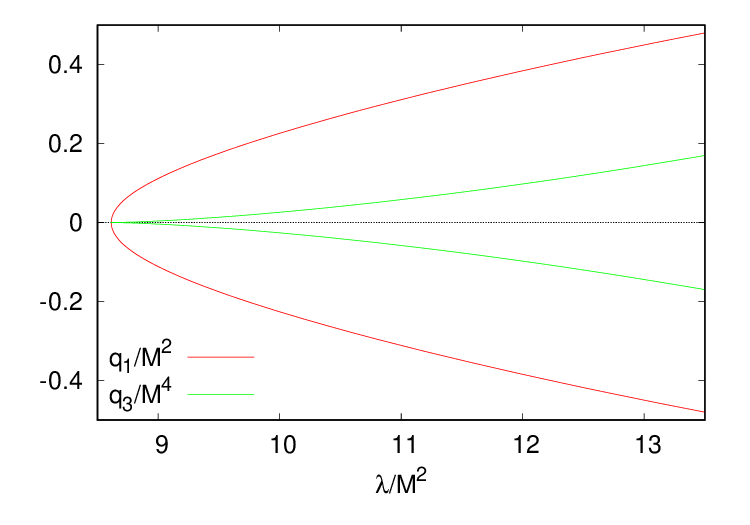}
}
\mbox{
(g)\hspace*{-0.5cm}\includegraphics[height=.22\textheight, angle =0]{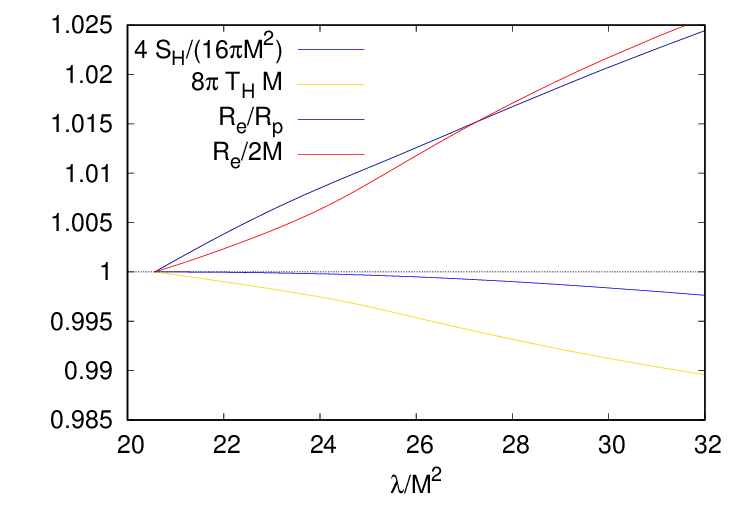}
(h)\hspace*{-0.5cm}\includegraphics[height=.22\textheight, angle =0]{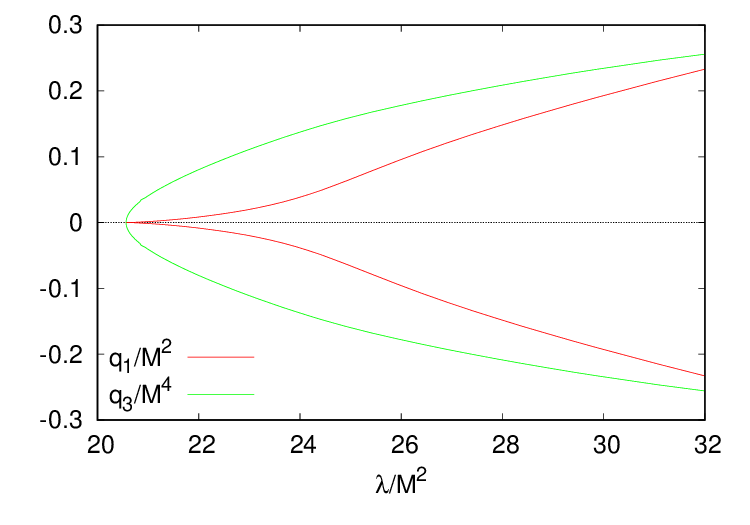}
}
\end{center}
\vspace{-0.5cm}
\caption{Static electric EvGB black holes.  
The left panels show scaled horizon and thermodynamic quantities, while the right panels show the corresponding scaled electric multipole moments. 
The rows correspond to $\ell=0$, $\ell=2$, $\ell=1$, and $\ell=3$, respectively.}
\label{fig:fig3}
\end{figure}

The physical properties are summarized in Fig.~\ref{fig:fig3}.  
For the spherically symmetric branch, panels (a) and (b) show the scaled entropy,
temperature, equatorial radius, and electric charge $q_0/M$.  
For the axisymmetric branches, panels (c)--(h) show the same type of scaled horizon and thermodynamic quantities together with the relevant electric multipole moments.
The branch with $\ell=2$ has even equatorial parity, while the branches with
$\ell=1$ and $\ell=3$ have odd equatorial parity.

The asymptotic electric vector field takes the form
\begin{equation}
V(r,\theta) \approx \sum_{\ell=0}^{\infty} q_\ell
\frac{P_\ell(\cos\theta)}{r^{\ell+1}} \ .
\label{A0asymp}
\end{equation}
The nonlinear branches are labelled by the angular index $\ell$ of the bifurcating perturbative mode.  
Away from the bifurcation point, however, the asymptotic field need not contain only the moment $q_\ell$.  
Equatorial symmetry selects either the even moments $q_0,q_2,\ldots$ or the odd moments $q_1,q_3,\ldots$.  
Thus an even branch may develop a moment such as $q_0$, while an odd branch may develop $q_1$ together with $q_3$.  
The right column of Fig.~\ref{fig:fig3} shows that the dominant moment grows rapidly as the coupling is increased from the bifurcation value, while further moments of the same parity, caused by the nonlinearities, grow more slowly.

\subsection{Nonlinear magnetic branches}\label{sec:magnetic}

Magnetic black holes are characterized by $A_\varphi\neq0$ and $A_t=0$.  
Analogous to the electric sector, they bifurcate from Schwarzschild at certain values $\lambda_\ell$ shown in Table~\ref{tab:lambda}.  
Their global structure is different, however.  
\jk{In terms of the dimensionless coupling used in the figures and tables, the magnetic branches exist only on finite intervals,
\begin{equation}
\frac{\lambda_{\rm cr,\ell}}{M^2}
\le
\frac{\lambda}{M^2}
\le
\frac{\lambda_\ell}{M^2} .
\end{equation}
They end at critical solutions whose first values are listed in  Table~\ref{tab:lambdacr}.
}
%The magnetic branches exist only on finite intervals,
%
%\begin{equation}
%\lambda_{{\rm cr},\ell} \leq \lambda \leq \lambda_\ell \ ,
%\end{equation}
%
%and end at critical solutions at $\lambda_{{\rm cr},\ell}$.  
%The first critical values are listed in Table~\ref{tab:lambdacr}.

\begin{table}[h!]
\begin{center}
\begin{tabular}{|c|c|c|}
\hline
$\ell$ & $\lambda_{{\rm cr},\ell}/M^2$  & $\lambda_\ell/M^2$ \\
\hline
0   &   2.911740 &  3.175894 \\
1   &   6.745346 &  7.297495 \\
2   &  11.884186 & 12.685844 \\
3   &  18.340467 & 19.383946 \\
\hline
\end{tabular}
\caption{Critical and bifurcation values for the magnetic branches with $\ell=0,\ldots,3$.}
\label{tab:lambdacr}
\end{center}
\end{table}

\begin{figure}[p!]
\begin{center}
\mbox{
(a)\hspace*{-0.5cm}\includegraphics[height=.22\textheight, angle =0]{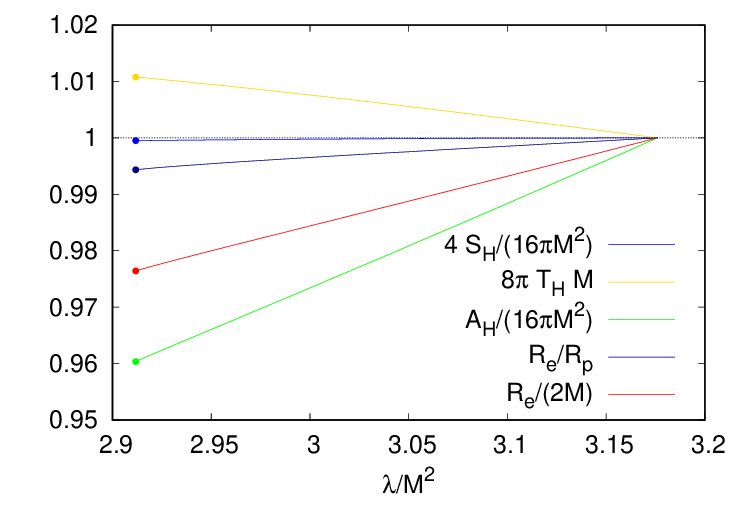}
(b)\hspace*{-0.5cm}\includegraphics[height=.22\textheight, angle =0]{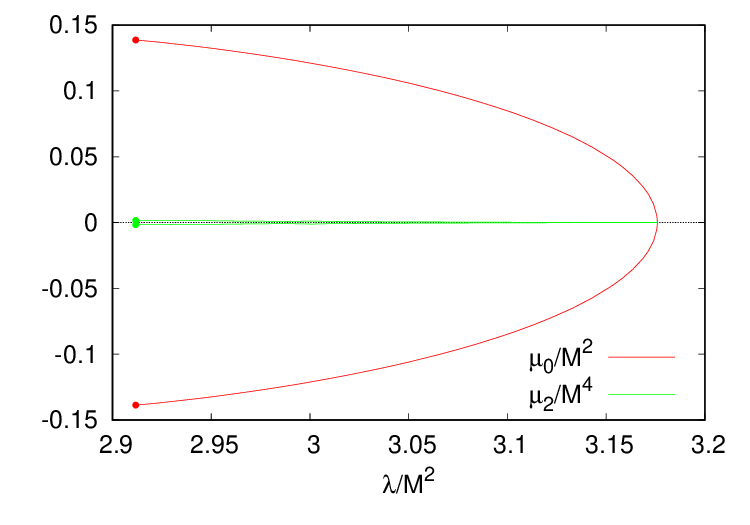}
}
\mbox{
(c)\hspace*{-0.5cm}\includegraphics[height=.22\textheight, angle =0]{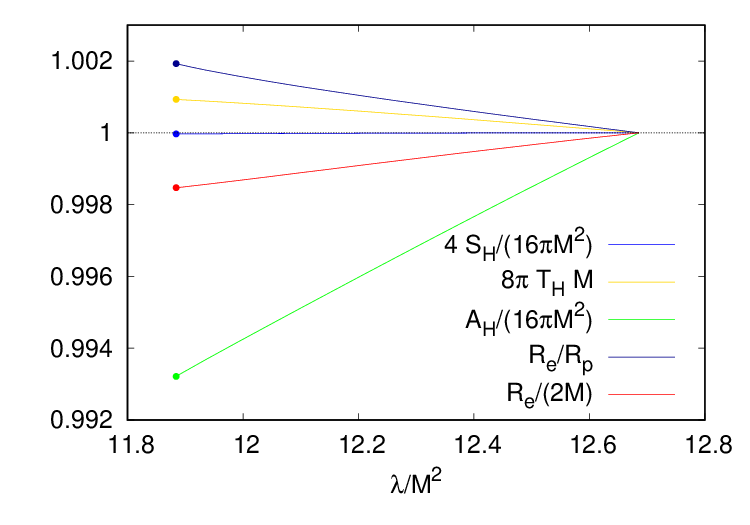}
(d)\hspace*{-0.5cm}\includegraphics[height=.22\textheight, angle =0]{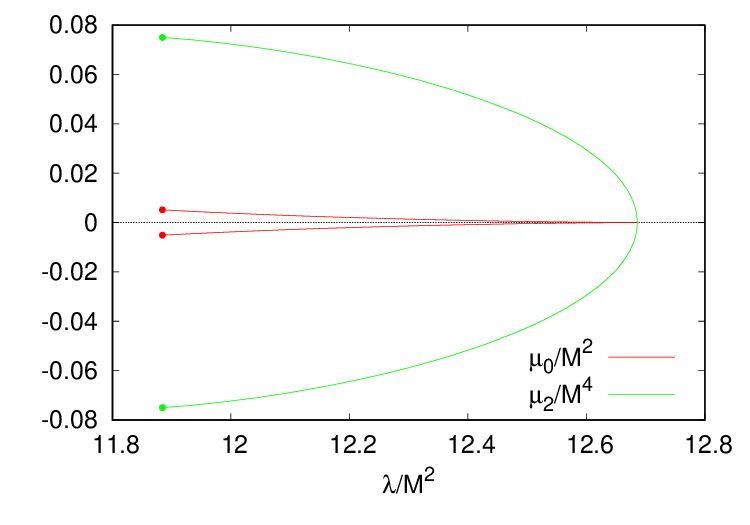}
}
\mbox{
(e)\hspace*{-0.5cm}\includegraphics[height=.22\textheight, angle =0]{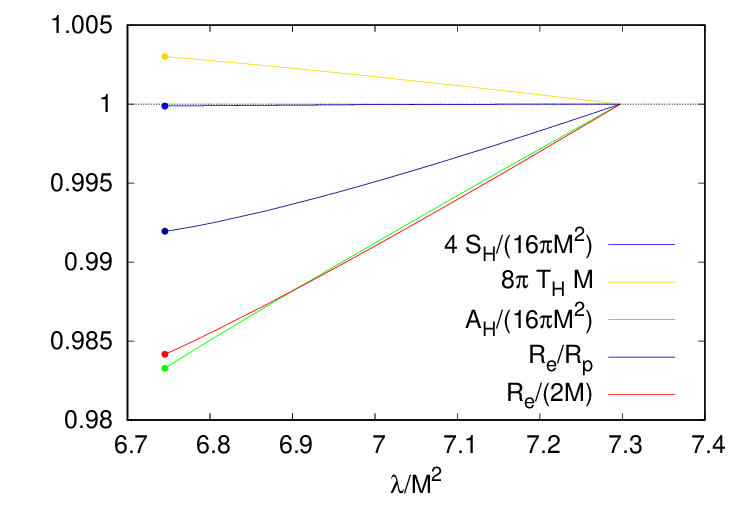}
(f)\hspace*{-0.5cm}\includegraphics[height=.22\textheight, angle =0]{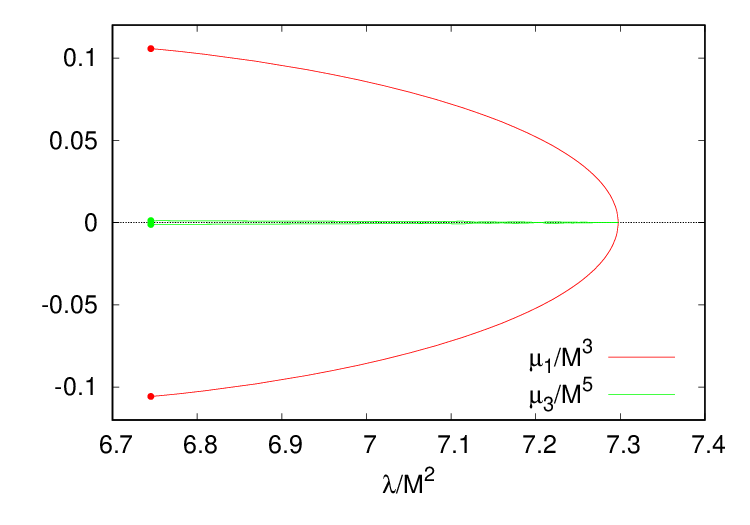}
}
\mbox{
(g)\hspace*{-0.5cm}\includegraphics[height=.22\textheight, angle =0]{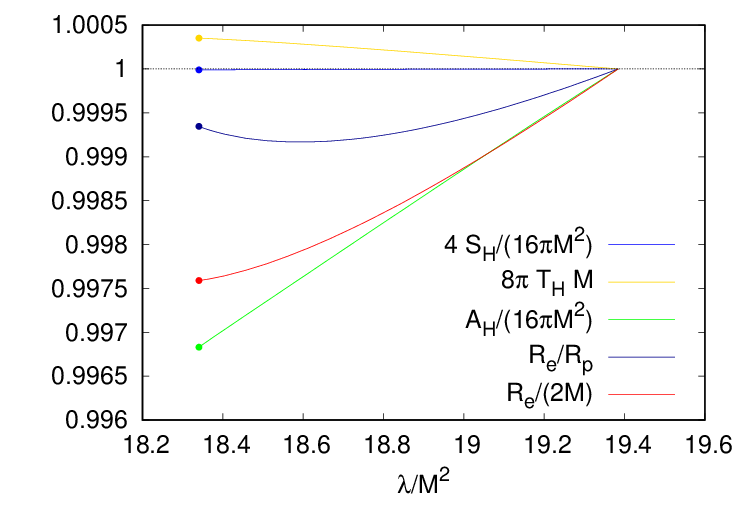}
(h)\hspace*{-0.5cm}\includegraphics[height=.22\textheight, angle =0]{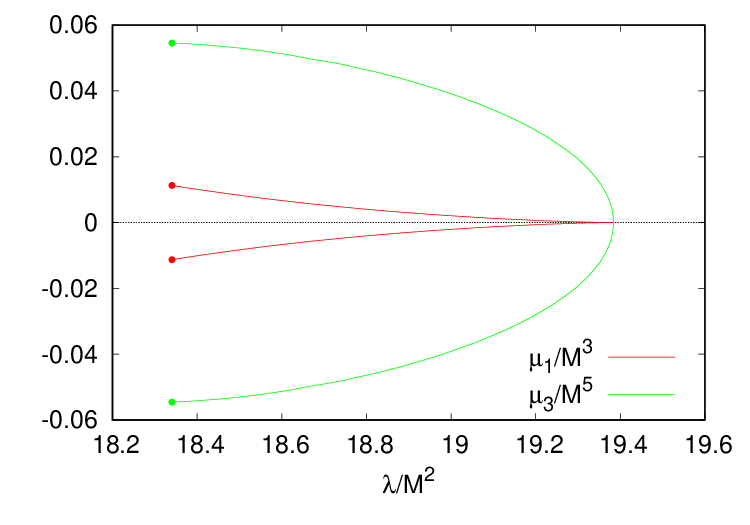}
}
\end{center}
\vspace{-0.5cm}
\caption{Static magnetic EvGB black holes.  
The left panels show scaled horizon and thermodynamic quantities, while the right panels show the corresponding scaled magnetic multipole moments.  
The rows correspond to $\ell=0$, $\ell=2$, $\ell=1$, and $\ell=3$, respectively.}
\label{fig:fig4}
\end{figure}

The physical properties are summarized in Fig.~\ref{fig:fig4}.  
The first row corresponds to the fundamental magnetic branch previously discussed in Ref.~\cite{Kleihaus:2026rev}.  
The further rows display the $\ell=2$, $\ell=1$, and $\ell=3$ branches.  
The left panels show the scaled entropy, Hawking temperature, horizon area, horizon deformation, and equatorial radius; the right panels show the scaled magnetic moments.

The asymptotic magnetic vector-field function is
\begin{equation}
H(r,\theta) \approx \sum_{\ell=0}^{\infty} \mu_\ell
\frac{\tilde{P}_\ell(\cos\theta)}{r^{\ell+1}} \ .
\label{Aphiasymp}
\end{equation}
Again, equatorial symmetry separates the nonlinear solutions into even and odd
multipole sectors. 
The scaled moments associated with the bifurcating mode grow rapidly as $\lambda/M^2$ is decreased from the bifurcation point toward the critical solution, while further allowed moments of the same parity grow more slowly.

\subsection{Comparison of electric and magnetic sequences}

Tables 1 and 2 reveal two systematic differences between the sectors. 
For every $\ell$ considered, the magnetic bifurcation occurs at a smaller value of $\lambda/M^2$ than the electric one. 
Moreover, the branches leave Schwarzschild in opposite directions in coupling space, and finite critical endpoints are found only in the magnetic sector.

The nonlinear multipole content also specifies how the branch label is to be interpreted.
The label $\ell$ refers to the perturbative mode from which the branch bifurcates, which is no longer the only asymptotic moment present in the fully nonlinear solution.  
Thus the multipole sequence consists of branches classified according to their bifurcation mode and parity.  
This is different from the radial excitations studied previously in Ref.~\cite{Kleihaus:2026rev}, where the angular structure was fixed while the radial function developed nodes.

\section{Conclusions}\label{sec:conclusions}

We have constructed branches of static multipolar vectorized black holes in EvGB theory with a quadratic coupling function.  
The angular multipole number $\ell$ labels the Schwarzschild perturbation that triggers the bifurcation.  
All branches constructed here have radial node number $n=0$.
The perturbative analysis yields discrete bifurcation values for the electric and magnetic sectors.  
The resulting nonlinear solution space consists of electric and magnetic multipolar branches with different domains of existence, while additional moments generated away from the bifurcation point remain restricted by equatorial parity.

We finally comment on the dynamical viability of such solutions. 
The present work is restricted to the construction and classification of stationary black-hole solutions and does not address their dynamical formation or stability. 
In related vectorization models, the tachyonic mechanism responsible for the growth of vector hair has been associated with ghost-like degrees of freedom or a loss of hyperbolicity of the vector-field equations \cite{Garcia-Saenz:2021uyv,Silva:2021jya,Clough:2022ygm,Coates:2022qia,Pizzuti:2023eyt,Chen:2024hkm}.  
More recently, a no-go argument has shown, within a broad class of vector-tensor theories and under its stated assumptions, that tachyonic vectorization around a hairless black hole is accompanied by ghost- or gradient-type instabilities \cite{Chiang:2026sow}. 
These results raise important concerns about the dynamical realization of spontaneous vectorization.  
However, whether their implications extend in full to the present quadratic EvGB model, and in particular to the nonlinear vectorized backgrounds constructed here, requires a dedicated analysis of time-dependent perturbations.

A remaining question is therefore not only whether the solution space of the present quadratic EvGB model can be enlarged, but also whether the multipolar branches found here are dynamically viable. 
It would likewise be interesting to determine whether comparable branches occur in other vector-tensor theories and whether the structural features identified here---the electric and magnetic sequences, their different domains of existence, and the parity selection of nonlinear multipole moments---persist more generally. 
The present solutions provide a reference for such studies, while their dynamical stability and possible physical relevance remain to be established.

\end{document}